\documentclass[tightenlines,twocolumn,english,preprintnumbers,amsmath,amssymb,nofootinbib,superscriptaddress]{revtex4-1}

\usepackage{appendix}
\usepackage{graphics,bm}
\usepackage{epsfig}
\usepackage{graphicx}
\usepackage{amsmath}
\usepackage{wrapfig}
\usepackage{color}

\newcommand{\beq}{\begin{equation}}
\newcommand{\eeq}{\end{equation}}
\newcommand{\bqa}{\begin{eqnarray}}
\newcommand{\eqa}{\end{eqnarray}}

\def\sumint{\hbox{$\sum$}\!\!\!\!\!\!\int}
\def\square{\vcenter{\vbox{\hrule height.4pt
          \hbox{\vrule width.4pt height4pt
          \kern4pt\vrule width.3pt}\hrule height.4pt}}}

\begin{document}

\title{Chiral density wave versus pion condensation
in the 1+1 dimensional NJL model}

\author{Prabal Adhikari}

\email{adhika1@stolaf.edu}
\affiliation{St. Olaf College, Physics Department, 1520 St. Olaf Avenue,
Northfield, MN 55057, USA}
\author{Jens O. Andersen}
\email{andersen@tf.phys.ntnu.no}
%\email{jens.andersen@ntnu.no}
\affiliation{Department of Physics, Faculty of Natural Sciences,NTNU, 
Norwegian University of Science and Technology, H{\o}gskoleringen 5,
N-7491 Trondheim, Norway}
\affiliation{Niels Bohr International Academy, 
Blegdamsvej 17, Copenhagen 2100, Denmark}
\date{\today}

\begin{abstract}
In this paper, we study the possibility of an inhomogeneous
quark condensate in the 1+1 dimensional Nambu-Jona-Lasinio
model in the large-$N_c$
limit at finite temperature $T$ and quark chemical potential $\mu$
using dimensional regularization.
The phase diagram in the $\mu$--$T$ plane is mapped out.
At zero temperature, an inhomogeneous phase with a chiral-density wave 
exists for $\mu>\mu_c$, where $\mu_c$ is a critical 
chemical potential.
Performing a Ginzburg-Landau analysis, we
show that in the chiral limit,
the tricritical point and the Lifschitz point coincide.
We also consider the competition between a chiral-density wave
and a constant pion condensate at finite isospin chemical potential $\mu_I$.
The phase diagram in the $\mu_I$--$\mu$ plane is mapped out and shows
a rich phase structure.

\end{abstract}
%\keywords{Dense QCD,chiral transition, }

\maketitle

\section{Introduction}
Confinement and the spontaneous breaking of chiral symmetry are
two of the most important properties of the 
vacuum of quantum chromodynamics (QCD).
The chiral condensate serves as an (approximate) order parameter for
the chiral transition:
At sufficiently high temperature or density, quarks are deconfined
and chiral symmetry is at least partly restored.
At asymptotically high temperature, QCD is a weakly interacting
quark-gluon plasma, and at asymptotically high density, QCD
is in the color-flavor locked phase and forms a color superconductor
\cite{rev1,rev2}.
At finite baryon chemical potential, lattice simulations
are difficult to perform
due to the infamous sign problem so one must use 
low-energy models for QCD.
At low temperature and high density, model calculations indicate that the
chiral transition is of first order.
This picture of a transition from a phase where chiral symmetry is
broken by a homogeneous chiral condensate to a phase where chiral symmetry
is (approximately) restored is probably too simplistic.
Model calculations also suggest that there is an inhomogeneous phase
in a relatively small region in the $\mu_B$--$T$ plane including
part of the $\mu_B$ axis.
%for a small range of (low) temperatures and (high) densities.
The idea of inhomogeneous phases at low temperature and high
density dates back to the 
work by  Fulde and Ferrell, and by
Larkin and Ovchinnikov in the context of superconductors
\cite{loff1,loff2}, density
waves in nuclear matter by Overhauser \cite{over}, and pion condensation by 
Migdal \cite{migdal}. 
In recent
years, inhomogeneous phases have been studied in, for
example, cold atomic gases \cite{baym}, color superconducting
phases \cite{incolor,anglo,heman}, quarkyonic phases \cite{robd,robd2}, 
pion condensates \cite{he1,he2}
as well as chiral condensates 
\cite{sadzi,nakano,nickel,nick2,bubsc,balli,abuki,braun,friman,dirk0,carigo3,dirk}, 
see Refs. \cite{suprev,buballarev} for recent
reviews.

In order to solve the problem of inhomogeneous phases in its
full generality, one must solve an infinite set
of coupled gap equations for the various Fourier modes.
This has not been done in three dimensions, but one hopes that a simple ansatz 
for the inhomogeneity will show many of the same features \cite{buballarev}.
Inhomogeneities that have been considered in 3+1 dimensions
are, for example,
one-dimensional modulations such as chiral-density waves and
soliton lattices. % \cite{nickel,bubsc}.

Field theories in 1+1 dimensions have been studied extensively over the
years as toy models for QCD since they have several important 
properties in common. For example, all 
Nambu-Jona-Lasinio (NJL) type models in 1+1 dimensions are
asymptotically free and show spontaneous breakdown of chiral symmetry
in the vacuum with a dynamically generated mass scale.
It should be pointed out, however, that in the 
NJL-type models in two dimensions
the breakdown of a continuous symmetry only takes place in the
large-$N_c$ limit, since the phase fluctuations that would otherwise
destroy a chiral condensate are of order 
$1/N_c$ \cite{mermin,cole}.
Although one is ultimately interested in 3+1 dimensions, the
models in 1+1 dimensions are ideal testing grounds for new
techniques. Calculations involving inhomogeneous phases can be found in 
Refs.~\cite{fullgn,full2,chiral11,1d1,1d0,massive,symren1,symren2,pionin, thieslast}.
One of the most important results in the past decade is
the construction of the exact phase diagram of the massive 
Gross-Neveu model in the large-$N_c$ limit \cite{fullgn,full2}.

In Ref. \cite{prabal}, we investigated systematically 
different regularization schemes in effective models with inhomogeneous 
phases. The vacuum energy of the NJL model in 1+1 was calculated
in the large-$N_c$ limit in the background of a chiral-density wave.
A naive application of for example momentum cutoff regularization
or dimensional regularization leads to an incorrect result
for the vacuum energy. The problem is that there is a residual dependence on
the wavevector $b$ in the limit where the magnitude $M$ 
goes to zero \cite{j+t,symren1}.
This unphysical behavior can be remedied by subtracting
the vacuum energy of a free Fermi gas
after having
performed a $b$-dependent unitary transformation
on the Hamiltonian.
We also showed that not all regulators are suited to perform
a Ginzburg-Landau (GL) analysis of the tricritical and Lifschitz points;
the proof of the equality of certain coefficients of the GL functional
sometimes involves integration by parts and requires that the
surface term vanishes. This is guaranteed if one uses 
dimensional regularization, but momentum cutoff regularization
fails
in certain cases, typically when the GL coefficients are divergent.

In this paper, we apply dimensional regularization 
and the techniques developed in Ref.~\cite{prabal} 
to calculate the free energy
to leading order in $N_c$ and map out the phase diagram in the $\mu$--$T$ plane
both in and away from the chiral limit. We also consider
the competition between a constant pion condensate and a chiral density
wave at finite isospin.
Our work is complementary to the study by Ebert et al \cite{pionin},
where the competition between a constant quark condensate
and an inhomogeneous pion condensate was studied at $T=0$ as a function
of $\mu$ and $\mu_I$.

The paper is organized as follows. In Sec. II, we briefly discuss
the NJL model in 1+1 dimensions and we derive the thermodynamic
potential at finite temperature and chemical potential
using dimensional regularization.
In Sec. III, we present the phase diagram and a Landau-Ginzburg
analysis of the critical and Lifschitz points. 
In Sec. IV, we discuss the competition between the chiral density
wave and a homogeneous pion condensate.
Finally, in Sec. V we summarize our results.
In the appendices, we provide the reader
with some calculational details of two sum-integrals that are needed
to locate the critical point and Lifschitz point.
We also discuss the vacuum energy 
in the special case of a finite pion condensate
and a vanishing chiral condensate.
%\begin{widetext}

\section{Lagrangian and Thermodynamic potential}
\begin{figure}[htb]
\end{figure}
%\subsection{Lagrangian and thermodynamic potential}
The Lagrangian of the NJL model in 1+1 dimensions is
\bqa\nonumber
{\cal L}&=&
\bar{\psi}\left[
i/\!\!\!\partial-m_0
+(\mu+\mbox{$1\over2$}\tau_3\mu_I
)\gamma^{0}
%-\sigma-i\gamma^{5}\pi_a\tau_a
\right]\psi
\\
&&+{G\over N_c}\left[
(\bar{\psi}\psi)^2+(\bar{\psi}i\gamma^5{\boldsymbol\tau}\psi)^2
\right]\;,
\label{lag1}
\eqa
where $N_c$ is the number of colors, $\tau_a$ are the three Pauli 
matrices ($a=1,2,3$) in isospin space,
$m_0$ is the current quark mass.
Moreover $\psi$ is a 
color $N_c$-plet, a two-component Dirac spinor, and a flavor doublet.
\bqa
\psi&=&
\left(
\begin{array}{c}
u\\
d
\end{array}\right)\;.
\eqa
The $\gamma$-matrices are $\gamma^0=\sigma_2$, $\gamma^1=i\sigma_1$, and
$\gamma^5=\gamma^0\gamma^1=\sigma_3$, where ${\sigma}_i$ are the
three Pauli matrices ($i=1,2,3$). 
Here
$\mu_B=3\mu={3\over2}(\mu_u+\mu_d)$ is the baryon chemical potential
and $\mu_I=\mu_u-\mu_d$ is the isospin chemical potential, 
where $\mu_f$ (with $f=u,d$) are the quark chemical potentials.
The Lagrangian (\ref{lag1}) is a generalization of the 
original Gross-Neveu model \cite{gross} which has a single quark flavor
and a single quark chemical potential.
The model (\ref{lag1})
has a global $SU(N_c)$ symmetry and for
$m_0=\mu_I=0$, it is also invariant under $SU_L(2)\times SU_R(2)$ 
transformations. For nonzero $m_0$ and $\mu_I=0$, 
the latter symmetry is reduced to the group $SU_I(2)$. 
For $m_0=0$ and nonzero $\mu_I$, it is reduced to 
$U_{I_3L}(1)\times U_{I_3R}(1)$.
Finally, for nonzero $m_0$ and $\mu_I$, the symmetry is 
reduced to $U_{I_3}(1)$.

We next introduce the bosonic fields $\sigma$ and $\pi_a$ via
\bqa
\sigma&=&-2{G\over N_c}\bar{\psi}\psi\;,
\\
\pi_a&=&-2{G\over N_c}\bar{\psi}i\gamma^5\tau_a\psi\;.
\eqa
The Lagrangian (\ref{lag1}) then becomes
\bqa\nonumber
{\cal L}&=&
\bar{\psi}\left[
i/\!\!\!\partial
-m_0
+(\mu
+\mbox{$1\over2$}\tau_3\mu_I
)\gamma^{0}-\sigma
-i\gamma^{5}\pi_a\tau_a
\right]\psi
\\ &&
-{N_c(\sigma^2+\pi_a^2)\over4G}\;.
\label{second}
\eqa
The chiral condensate we choose is a chiral-density wave
of the form 
\bqa
\langle\sigma\rangle&=&M \cos(2bz)-m_0
\;,
\label{back0}
\\
\langle\pi_3\rangle&=&M\sin(2bz)\;,
\label{back}
\eqa
where $b$ is a wavevector. For $b=0$, it reduces to the 
standard homogeneous condensate.
With a nonzero isospin chemical potential, there is 
also the possibilty of a pion condensate $\Delta$. 
%{\color{blue}The charged pions will condense once their mass exceeds the 
%isospin chemical potential, $m_{\pi}\geq\mu_I$. In this phase
%it is a massless Goldstone boson associated with the breaking
%of $U_{I_3}(1)$ symmetry.}
For simplicity, we take this to be homogeneous
\bqa
\langle\pi_1\rangle&=&\Delta\;.
\label{back2}
\eqa
The last term in Eq. (\ref{second}) is denoted by $-V_0$, where
$V_0$ is the tree-level potential.
Inserting Eqs. (\ref{back0})--(\ref{back2}) into 
$V_0$ and averaging over the spatial extent
$L$ of the system, we obtain 
for $L\rightarrow\infty$
\bqa
V_0&=&N_c{M^2+m_0^2-2Mm_0\delta_{b,0}+\Delta^2\over4G}\;.
\eqa
In the homogeneous case, the tree-level potential reduces to the
standard expression $V_0=N_c{(M-m_0)^2+\Delta^2\over4G}$.

The Dirac operator $D$ can be written as 
\bqa\nonumber
D&=&\bar{\psi}\big[
i/\!\!\!\partial
+(\mu+
\mbox{$1\over2$}\tau_3\mu_I)\gamma^0
%i\gamma^{\mu}\partial_{\mu}
-Me^{2i\gamma^5\tau_3bz}
\\&&
-i\gamma^5\tau_1\Delta
\big]\psi\;.
\label{dirac}
\eqa
We next redefine the quark fields, $\psi\rightarrow e^{-i\gamma^5\tau_3bz}\psi$
and $\bar{\psi}\rightarrow\bar{\psi}e^{-i\gamma^5\tau_3bz}$, which
corresponds to a unitary transformation of the 
Dirac Hamiltonian, % of the system,
${\cal H}\rightarrow{\cal H}^{\prime}
=e^{i\gamma^5\tau_3bz}{\cal H}e^{-i\gamma^5\tau_3bz}$.
The Dirac operator then reads
\bqa
D&=&
\left[i/\!\!\!\partial
%i\gamma^{\mu}\partial_{\mu}
+(\mu%\gamma^0
%+(b+\mbox{$1\over2$}\mu_I)
+b^{\prime}
\tau_3)\gamma^0
-M-i\gamma^5\tau_1\Delta
\right]\;,
%\label{dirac}
\eqa
where $b^{\prime}=(b+\mbox{$1\over2$}\mu_I)$.
Going to momentum space, Eq. (\ref{dirac}) can be written as
\bqa
D&=&\left[/\!\!\!p
%i\gamma^{\mu}p_{\mu}
+(\mu%\gamma^0
%+(b+\mbox{$1\over2$}\mu_I)
+b^{\prime}\tau_3)\gamma^0
-M-i\gamma^5\tau_1\Delta
\right]\;.
\label{shows}
\eqa
Eq. (\ref{shows}) shows that the effective chemical potential for the
$u$-quarks is $\mu+b^{\prime}=\mu_u+b$, while for the $d$-quarks, it is 
$\mu-b^{\prime}=\mu_d-b$.
It is now straightforward to
derive the fermionic spectrum in the background 
(\ref{back})--(\ref{back2}).
It is given by the zeros of the Dirac determinant 
and reads \cite{massive,symren1}	
\bqa
p_{0u}&=&E_{\Delta}^{-}-\mu\;,
\hspace{0.65cm}
p_{0d}=E_{\Delta}^{+}-\mu\;,
\\
p_{0\bar{u}}&=&-(E_{\Delta}^{+}+\mu)\;,
\hspace{0.1cm}
p_{0\bar{d}}=-(E_{\Delta}^{-}+\mu)\;,
\eqa
where
\bqa
E_{\Delta}^{\pm}&=&\sqrt{E_{\pm}^2+\Delta^2}\;,
\hspace{0.1cm}E_{\pm}=\sqrt{p^2+M^2}\pm b^{\prime}\;.
\eqa
We note that the spectrum depends on the isospin chemical potential
$\mu_I$ via $b^{\prime}$.

Going to Euclidean space,
the one-loop contribution to the thermodynamic potential 
is given by
\bqa
V_1&=&-N_c\sumint_{\{P\}}\log\left[P_0^2+(E_{\Delta}^{\pm})^2\right]\;,
\eqa
where the sum-integral is defined in Eq. (\ref{deffie})
and a sum over $\pm$ is implied.
Summing over the Matsubara frequencies, we can write
\bqa\nonumber
V_1&=&-N_c\int_p\left\{
E_{\Delta}^{\pm}+T\log\left[1+e^{-\beta(E_{\Delta}^{\pm}-\mu)}\right]
\right. \\ &&\left.
+T\log\left[1+e^{-\beta(E_{\Delta}^{\pm}+\mu)}\right]\right\}\;,
\label{total}
\eqa
where the integral is defined in Eq. (\ref{def2}).
The first term in Eq. (\ref{total}) is ultraviolet divergent and requires
regularization. The two contributions from $E_{\Delta}^{\pm}$
to this term are denoted by $V_{\pm}^{\rm vac}$.
The second and third terms which depend on the temperature
and the chemical potential are finite.

After integrating over angles and changing variables, $u=\sqrt{p^2+M^2}$, we
can write
\bqa\nonumber
V_{\pm}^{\rm vac}&=&
-{N_c(e^{\gamma_E}\Lambda^2)^{\epsilon}\over\sqrt{\pi}\Gamma({1\over2}-\epsilon)}
\int_M^{\infty}\sqrt{(u\pm b^{\prime})^2+\Delta^2}
\\&&
\hspace{2.2cm}\times
{udu\over(u^2-M^2)^{{1\over2}+\epsilon}}\;.
\eqa
We cannot calculate analytically the vacuum energy for nonzero $\Delta$.
In order to isolate the divergences, we expand the dispersion relations
around $\Delta=0$
and find appropriate subtraction terms. 
%After integration over angles, we can write
We can then write
\bqa
V_{\pm}^{\rm vac}&=&V_{\rm div\pm}^{\rm vac}+V_{\rm fin\pm}^{\rm vac}\;,
\eqa
where
\begin{widetext}
\bqa
\label{divida}
V_{\rm div\pm}^{\rm vac}&=&
-{N_c(e^{\gamma_E}\Lambda^2)^{\epsilon}\over\sqrt{\pi}\Gamma({1\over2}-\epsilon)}
\left[\int_M^{\infty}
|u\pm b^{\prime}|
+{\Delta^2\over2u}\right]{udu\over(u^2-M^2)^{{1\over2}+\epsilon}}
\;,
\\
V_{\rm fin\pm}^{\rm vac}
&=&
-{N_c(e^{\gamma_E}\Lambda^2)^{\epsilon}\over\sqrt{\pi}\Gamma({1\over2}-\epsilon)}
\int_M^{\infty}
\left[E_{\Delta}^{\pm}-|u\pm b^{\prime}|-{\Delta^2\over2u}\right]
{udu\over(u^2-M^2)^{{1\over2}+\epsilon}}\;.
\label{fin2}
\eqa
\end{widetext}
We denote the sum of the two terms in (\ref{fin2}) by 
$V_{\rm fin}^{\rm vac}$.
Note that $V_{\rm fin\pm}^{\rm vac}=0$ for $\Delta=0$.
In the chiral limit, the solutions to the gap equations 
${\partial V\over\partial M}={\partial V\over\partial\Delta}=0$
(with $V=V_0+V_1$)
are $M\neq0$ and $\Delta=0$ or $M=0$ and $\Delta\neq0$. In the latter case,
Eqs. (\ref{divida}) and (\ref{fin2}) are infrared divergent. The IR divergences of  (\ref{divida}) cancel
against those of (\ref{fin2}). However, they must be regulated
separately, which is inconvenient. In Appendix A, we discuss
this case.

$V_{\rm div\pm}^{\rm vac}$ can now be calculated using dimensional regularization
and the result is 
\bqa\nonumber
V_{\rm div+}^{\rm vac}&=&{N_c\over4\pi}\left(
{e^{\gamma_E}\Lambda^2\over M^2}\right)^{\epsilon}
\left[M^2\Gamma(-1+\epsilon)-\Delta^2\Gamma(\epsilon)\right]\;,
\\
&& \\
V_{\rm div-}^{\rm vac}&=&V_{\rm div+}^{\rm vac}
+\theta(b^{\prime}-M)f(M,b^{\prime})\;,
\label{uss}
\eqa
\begin{widetext}
where the function $f(M,b^{\prime})$ is defined by
\bqa%\nonumber
f(M,b^{\prime})&=&
-{N_c\over\pi}\left[b^{\prime}\sqrt{b^{\prime2}-M^2}-
M^2\log{b^{\prime}+\sqrt{b^{\prime2}-M^2}\over M}\right]\;.%\theta(b-M)
\eqa
\end{widetext}
The contribution $V_{\rm div+}^{\rm vac}$ to the vacuum energy is
independent of $b$, while
the extra term $f(M,b^{\prime})$ in $V_{\rm div-}^{\rm vac}$
%in Eq. (\ref{divida}), 
arises from the integral
$\int_M^{\infty}|u-b^{\prime}|$ 
where one must distinguish between 
$u<b^{\prime}$ and $u>b^{\prime}$.

Expanding $V_{\rm div}^{\rm vac}=V_{\rm div+}^{\rm vac}+V_{\rm div-}^{\rm vac}$
in powers of $\epsilon$, we find
\bqa\nonumber
V_{\rm div}&=&-{N_c\over2\pi}\left(
{\Lambda^2\over M^2}\right)^{\epsilon}
\left[\left({1\over\epsilon}+1\right)M^2
+{1\over\epsilon}\Delta^2
\right]
\\ &&
+\theta(b^{\prime}-M)f(M,b^{\prime})
\label{difs}
\;.
\eqa
Eq. (\ref{difs}) contains poles in $\epsilon$ that 
are removed by the renormalization of the 
the fermion mass $m_0$ and the (inverse) coupling constant $G$ by 
making the substitutions
$m_0\rightarrow Z_{m_0}m_0$
and ${1\over G}\rightarrow Z_{G^{-1}}{1\over G}$, where
\bqa
\label{sub1}
Z_{m_0}&=&\left[1+{2G\over\pi\epsilon}\right]^{-1}\;,
%{1\over G}&\rightarrow&{1\over G}\left[1+{2G\over\pi\epsilon}\right]\;,
\\
Z_{G^{-1}}&=&
\left[1+{2G\over\pi\epsilon}\right]\;.
%m_0&\rightarrow&m_0\left[1+{2G\over\pi\epsilon}\right]^{-1}\;.
\label{sub2}
\eqa 
Note that $Z_{G^{-1}}=Z_G^{-1}$ and that 
the ratio 
${m_0\over G}$ is the same for bare and renormalized
quantities since $Z_{m_0}Z_G^{-1}=1$.
After renormalization, making the substitutions Eqs.
(\ref{sub1}) and (\ref{sub2}), 
the vacuum energy $V=V_0+V_1$ becomes
\begin{widetext}
\bqa\nonumber
V&=&N_c{(M^2+m_0^2-2Mm_0\delta_{b^{},0})+\Delta^2\over4G}
-{N_cM^2\over2\pi}\left[\log{\Lambda^2\over M^2}+1\right]
-{N_c\Delta^2\over2\pi}\log{\Lambda^2\over M^2}
+V_{\rm fin}^{\rm vac}
%-{N_c\over2}mM
\\ &&
+\theta(b^{\prime}-M)f(M,b^{\prime})\;.
%-\theta(b^{\prime}-m_0)f(m_0,b^{\prime})
%+\theta(\mbox{$1\over2$}\mu_I-m_0)f(m_0,\mbox{$1\over2$}\mu_I)\;,
\label{effpot000}
\eqa
%where we have added a vacuum counterterm
%$\Delta{V}=-{N_cm_0^2\over2}{1\over2G+\pi\epsilon}$.
\end{widetext}
%$V_{\rm fin}^{\rm vac}=V_{\rm fin+}^{\rm vac}+V_{\rm fin-}^{\rm vac}$.
Due to the term $b^{\prime}\sqrt{b^{\prime2}-M^2}$ in the 
function $f(M,b^{\prime})$, the 
vacuum energy is unbounded from below. 
For $m_0=0$, and $\Delta=M=0$, $V=-{N_c\over\pi}b^{\prime2}$ 
and depends on $b^{\prime}$, which is unphysical
(the special case $M=0$ and $\Delta\neq0$ is discussed
in Appendix \ref{app2}.). 
The same problem occurs if one uses a momentum 
cutoff and in \cite{symren1} 
it was suggested to subtract the term
$V_{\rm sub}=-{N_c\over\pi}b^{\prime2}+{N_c\over4\pi}\mu_I^{2}$, where the
latter is necessary to ensure to correct expression of the vacuum energy
in the limit 
$b\rightarrow0$.\footnote{If one uses an energy cutoff \cite{symren1}, 
there is no spurious $b$-dependence, but one still has to subtract
a term $V_{\rm sub}={N_c\over4\pi}\mu_I^{2}$.}

As explained in the introduction, we suggest to subtract the vacuum energy for the 
system of a free Fermi gas (after a unitary transformation)
in order to obtain a result that is
independent of $b$ in the limit $M\rightarrow0$.
Thus we subtract the term
\bqa\nonumber
V_{\rm sub}&=&
%-{N_c\over2\p\left(
-{N_cm_0^2\over2\pi}\left[\log{\Lambda^2\over m_0^2}+1\right]
%{\Lambda^2\over m_0^2}\right)^{\epsilon}
%m_0^2\left({1\over\epsilon}+1\right)
+\theta(b^{\prime}-m_0)f(m_0,b^{\prime})
\\ &&
-\theta(\mbox{$1\over2$}\mu_I-m_0)f(m_0,b^{\prime})
\;.
\label{subbi}
\eqa
Eq. (\ref{subbi}) then
reduces to 
$V_{\rm sub}=-{N_c\over\pi}b^{\prime2}+{N_c\over4\pi}\mu_I^{2}$
for $m_0=0$. Moreover, the first term in Eq. (\ref{subbi}) is
independent of the parameter $b$ and the chemical potential
$\mu_I$ and can therefore be omitted. The final result for the vacuum energy
is therefore
\begin{widetext}
\bqa\nonumber
V&=&N_c{(M^2+m_0^2-2Mm_0\delta_{b^{},0})+\Delta^2\over4G}
-{N_cM^2\over2\pi}\left[\log{\Lambda^2\over M^2}+1\right]
-{N_c\Delta^2\over2\pi}\log{\Lambda^2\over M^2}
+V_{\rm fin}^{\rm vac}
%-{N_c\over2}mM
\\ &&
+\theta(b^{\prime}-M)f(M,b^{\prime})
-\theta(b^{\prime}-m_0)f(m_0,b^{\prime})
+\theta(\mbox{$1\over2$}\mu_I-m_0)f(m_0,\mbox{$1\over2$}\mu_I)\;.
\label{effpot0}
\eqa
\end{widetext}
The finite-temperature term is the second and third terms from
(\ref{total}),
\bqa\nonumber
V^T_1&=&
-{N_cT\over\pi}\int_0^{\infty}\bigg\{
\log\left[1+e^{-\beta(E_{\Delta}^{\pm}-\mu)}\right]
\\ 
&&
\hspace{1.3cm}
+\log\left[1+e^{-\beta(E_{\Delta}^{\pm}+\mu)}\right]
\bigg\}\,dp\;.
%\\ &&
\label{fint}
\eqa
The complete free energy in the large-$N_c$ limit is then given by
the sum of Eq. (\ref{effpot0}) and Eq. (\ref{fint}).
In contrast to 3+1 dimensions, we have no experimental input that
allows us to determine the constituent quark mass $m_0$
appearing in the expression for the free energy.
Following Ref. \cite{massive}, we demand that the ratio of the dynamical
quark mass $M$ and the pion mass $m_{\pi}$ be the same as in three dimensions
for $\mu=\mu_I=0$. 
Choosing the values $M=350$ MeV and $m_{\pi}=140$ MeV, one finds
a ratio ${M\over m_{\pi}}={5\over2}$. Numerically, this corresponds to
values $m_0=0.05M_0$, $M=1.04 M_0$, and $m_{\pi}=0.42M_0$, where
$M_0$ is the dynamical quark mass for $m_0=0$. 
Introducing the dimensionless $\alpha=\pi{m_0\over M_0}$, this corresponds
to $\alpha=0.17$. In the remainder of the paper. we use this
value for $\alpha$. Moreover, since all contributions to the effective
potential and gap equations 
are proportional to $N_c$, we
omit this factor in all the numerical work.

We close this section by discussing the running parameters in the model
and the solution in the vacuum. 
The coupling constant $G$ and the mass parameter $m_0$ satisfy
the renormalization group equations
\bqa
\label{rg1}
\Lambda{dG\over d\Lambda}&=&-{4G^2\over\pi}\;,
\\
\Lambda{dm_0\over d\Lambda}&=&-{4m_0G\over\pi}\;.
\label{rg2}
\eqa
%These equations show that the vacuum part of the effective potential
%Eq. (\ref{effpot0}) is renormalization group invariant.
The solutions are 
\bqa
G(\Lambda)
&=&
{G(\Lambda_0)\over1+{4\over\pi}G(\Lambda_0)\log{\Lambda\over\Lambda_0}}\;,
\\
m_0(\Lambda)&=&
{m_0(\Lambda_0)\over G(\Lambda_0)}
G(\Lambda)\;,
\eqa
where $\Lambda_0$ is some reference scale. 
These equations show that the ratio
${m_0\over G}$ is independent of the scale $\Lambda$.
We also note that $G(\Lambda)$ decreases with $\Lambda$ showing
that the model is asymptotically free.

In the vacuum phase, we have $\Delta=b=0$, and in the chiral limit, 
the solutions $M_0$ to the gap equation ${dV\over dM}=0$
are either $M_0=0$ or
\bqa
M_0&=&\Lambda e^{-{\pi\over4G}}\;.
\label{nona}
\eqa
Using Eq. (\ref{rg1}), it is straightforward to verify that
$M_0$ is independent of the renormalization scale $\Lambda$.
The nonanalytic behavior of $M_0$ as a function of $G$
shows that the result is nonperturbative. Using for example
the two-particle irreducible action formalism, it can be shown
that this results corresponds the summation of the daisy and 
superdaisy graphs from all orders of perturbation
theory \cite{gert,jens}.
Using Eq. (\ref{nona}),
we can trade the scale $\Lambda$ for the scale $M_0$, which gives
\bqa
V&=&
-{N_cM^2\over2\pi}
\left[
\log\left({M_0^2\over M^2}\right)+1
\right]\;,
\label{vacci}
\eqa
%\end{widetext}
in agreement with Ebert et al \cite{symren1}. 
It is easy to see that the global minimum of $V$ is at $M=M_0$.
In the remainder of this paper, we express all dimensionful quantities
in appropriate powers of the dynamically generated mass scale $M_0$.

\section{Chiral-density wave and no pion condensate ($\Delta=0$)} 
In the absence of a pion condensate, the vacuum energy (\ref{effpot0})
reduces to
\begin{widetext}
\bqa\nonumber
V&=&N_c{(M^2+m_0^2-2Mm_0\delta_{b,0})\over4G}
-{N_cM^2\over2\pi}
\left[\log{\Lambda^2\over M^2}+1\right]
+\theta(b^{\prime}-M)f(M,b^{\prime})
\\ &&
-\theta(b^{\prime}-m_0)f(m_0,b^{\prime})
+\theta(\mbox{$1\over2$}\mu_I-m_0)f(m_0,\mbox{$1\over2$}\mu_I)\;,
\label{effpot01}
\eqa
\end{widetext}
where we have used that $V_{\rm fin}^{\rm vac}=0$ for $\Delta=0$.
The finite-temperature term is given by Eq. (\ref{fint})
evaluated for $\Delta=0$.
\subsection{Zero temperature}
In the limit $T\rightarrow0$ and for vanishing pion condensate, $\Delta=0$, 
one can obtain analytic results for the
density-dependent contributions to the effective potential given 
by Eq. (\ref{fint}).
The contributions from the first term in Eq. (\ref{fint})
are denoted by $V_{\pm}^{\rm med}$ and read
%. After integrating over angles, we find
\bqa
V_{\pm}^{\rm med}&=&-{N_c\over\pi}\int_0^{\infty}(\mu-E_{\pm})\theta(\mu-E_{\pm})
\,dp\;.
\label{denni}
\eqa
The contributions from the second term in Eq. (\ref{fint})
vanish for $\mu>0$
and vice versa for $\mu<0$. Without loss of generality
we take $\mu>0$ in the remainder.
The contribution $V_+^{\rm med}$
is straightforward to compute. 
After changing variables $u=\sqrt{p^2+M^2}$ and noting that the
upper limit is $u_f=\mu-b^{\prime}$ due to the step function,
we find
\begin{widetext}
\bqa\nonumber
V_+^{\rm med}&=&
-{N_c\over\pi}\int_0^{\infty}(\mu-E_+)\theta(\mu-E_+)\,dp
\\ \nonumber
&=&-{N_c\over\pi}\int_M^{u_f}(\mu-u-b^{\prime})
{u\,du\over\sqrt{u^2-M^2}}
\\ &=&
-{N_c\over2\pi}
\left[
(\mu-b^{\prime})\sqrt{(\mu-b^{\prime})^2-M^2}
-M^2\log{\mu-b^{\prime}+\sqrt{(\mu-b^{\prime})^2-M^2}\over M}
\right]\theta(\mu-b^{\prime}-M)\;.
\label{va}
\eqa
%\end{widetext}
We next consider the contribution $V_-^{\rm med}$, which is given by
%\bqa
%V_-^{\rm med}&=&
%-{1\over\pi}\int_0^{\infty}(\mu-E_-)\theta(\mu-E_-)\;.
%\eqa
\bqa
V_-^{\rm med}&=&
-{N_c\over\pi}\int_0^{\infty}(\mu-E_-)\theta(\mu-E_-)\,dp\;.
\eqa
Here we must distinguish between several cases.

\begin{enumerate}
\item
$M>b^{\prime}$. 
The dispersion relation is shown in the left panel of Fig. \ref{density1}.
%In this case $E_-=\sqrt{p^2+M^2}-b^{\prime}$. 
In this case, the integration is from $p=0$ to 
$p_C=p_f=\sqrt{(\mu+b^{\prime})^2-M^2}$ or $u=M$ to
to $u=u_f=\mu+b^{\prime}$,
\bqa
V_-^{\rm med}&=&-{N_c\over\pi}\int_M^{u_f}(\mu-u+b^{\prime})
{u\,du\over\sqrt{u^2-M^2}}
\;,
\eqa
where %$u_f=\mu+b^{\prime}$ and 
$\mu>M-b^{\prime}$. This yields
\bqa
V_-^{\rm med}&=&-{N_c\over2\pi}
\left[
(\mu+b^{\prime})\sqrt{(\mu+b^{\prime})^2-M^2}
-M^2\log{\mu+b^{\prime}+\sqrt{(\mu+b^{\prime})^2-M^2}\over M}\right]
\theta(\mu+b^{\prime}-M)
\;.
\eqa
This contribution is obtained from (\ref{va}) by the substitution
$b^{\prime}\rightarrow-b^{\prime}$.
\item $b^{\prime}>M$. 
The dispersion relation is shown in the right panel of Fig. \ref{density1}
(blue curve).
In this case
$E_-=b^{\prime}-\sqrt{p^2+M^2}$ for $u<b^{\prime}$ and
$E_-=\sqrt{p^2+M^2}-b^{\prime}$ for $u>b^{\prime}$. 
\begin{enumerate}
\item If $\mu>b^{\prime}-M$, the integration is from $p=0$ to 
$p_C=p_f=\sqrt{(b^{\prime}+\mu)^2-M^2}$ or
$u=M$ to $u=u_f=\mu+b^{\prime}$. The green horizontal line indicates the
value of the chemical potential and the intersection with the dispersion
relation gives the upper limit of integration.
This yields
\bqa\nonumber
V_-^{\rm med}&=&-{N_c\over\pi}
\bigg[
{1\over2}(\mu+b^{\prime})\sqrt{(\mu+b^{\prime})^2-M^2}
-b^{\prime}\sqrt{b^{\prime2}-M^2}
+M^2\log{b^{\prime}+\sqrt{b^{\prime2}-M^2}\over M}
\\ &&
%\times
-{1\over2}M^2\log{\mu+b^{\prime}+\sqrt{(\mu+b^{\prime})^2-M^2}\over M}
\bigg]
\theta(\mu-b^{\prime}+M)\;.
\eqa
\item If $\mu<b^{\prime}-M$, the integration is from 
$p_A=\sqrt{(b^{\prime}-\mu)^2-M^2}$ to $p_B=p_f=\sqrt{(b^{\prime}+\mu)^2-M^2}$
or $u=b^{\prime}-\mu$ to $u=b^{\prime}+\mu$. The value of the chemical potential is
indicated by the orange line and the intersection with the dispersion relation
gives the upper and lower limits of integration.
This gives
\bqa\nonumber
V_-^{\rm med}
&=&-{N_c\over\pi}
\left[
{1\over2}(b^{\prime}+\mu)\sqrt{(b^{\prime}+\mu)^2-M^2}+
{1\over2}(b^{\prime}-\mu)\sqrt{(b^{\prime}-\mu)^2-M^2}
-b^{\prime}\sqrt{b^{\prime2}-M^2}
\right.\\ &&\left.
-{1\over2}M^2
\log{b^{\prime}+\mu+\sqrt{(b^{\prime}+\mu)^2-M^2}
\over b^{\prime}+\sqrt{b^{\prime2}-M^2}}
-{1\over2}M^2\log{b^{\prime}-\mu+\sqrt{(b^{\prime}-\mu)^2-M^2}
\over b^{\prime}+\sqrt{b^{\prime2}-M^2}}
\right]\theta(b^{\prime}-\mu-M)\;.
\eqa
\end{enumerate}
\end{enumerate}
Combining the different cases discussed above, 
the result for the 
full effective potential in the large-$N_c$ limit can be written as
\bqa\nonumber
V^{\rm }&=&
N_c{(M^2+m_0^2-2Mm_0\delta_{b,0})\over4G}
-{N_cM^2\over2\pi}\left[\log{M^2\over M_0^2}+1\right]
-\theta(b^{\prime}-m_0)f(m_0,b^{\prime})
+\theta(\mbox{$1\over2$}\mu_I-m_0)f(m_0,\mbox{$1\over2$}\mu_I)\
%+{N_cb^{\prime2}\over\pi}
\\ && \nonumber
-{N_c\over2\pi}\left[(\mu+b^{\prime})\sqrt{(\mu+b^{\prime})^2-M^2}
-M^2\log{\mu+b^{\prime}+\sqrt{(\mu+b^{\prime})^2-M^2}\over M}
\right]\theta(\mu+b^{\prime}-M)
\\
&&-{N_c\over2\pi}\left[|\mu-b^{\prime}|\sqrt{(\mu-b^{\prime})^2-M^2}
-M^2\log{|\mu-b^{\prime}|+\sqrt{(\mu-b^{\prime})^2-M^2}\over M}
\right]\theta(|\mu-b^{\prime}|-M)
\;.
\label{fullkli}
\eqa

\begin{figure}[htb]
\begin{center}
\includegraphics[width=0.4\textwidth]{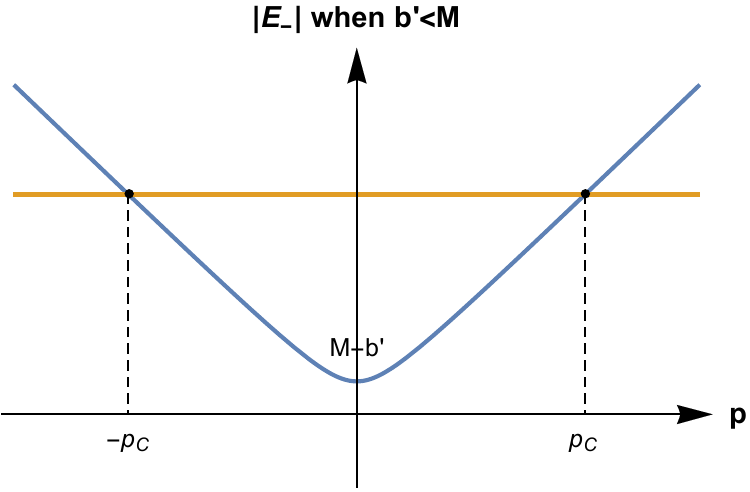}
\hspace{0.4cm}
\includegraphics[width=0.4\textwidth]{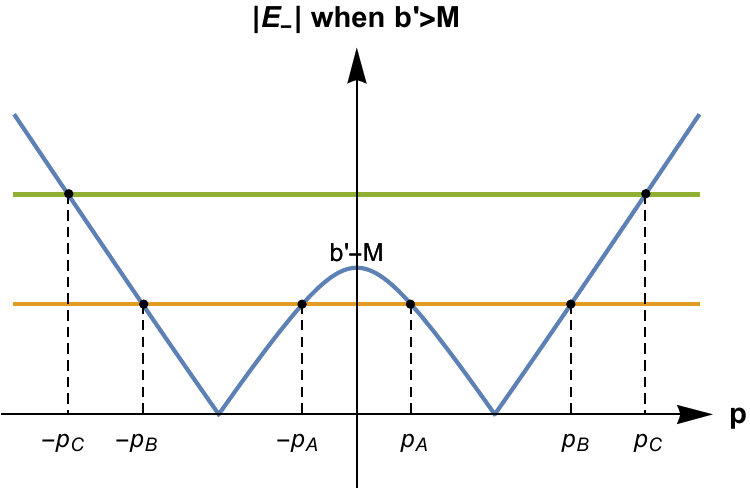}
\end{center}
\caption{(Color online) Dispersion relation $E_-$ for $\Delta=0$
(blue curve) for $b^{\prime}<M$ (left panel)
and for $b^{\prime}>M$ (right panel). The horizontal green line is for the case
$\mu>b^{\prime}-M$ 
and the horizontal orange line is for the case $\mu<b^{\prime}-M$.
See main text for discussion of the 
regions of integration in the different cases.}
\label{density1}
\end{figure}
\end{widetext}
In Fig. \ref{larger}, we show the magnitude $M$ (blue solid line)
and the wavevector
$b$ (red dashed line) both normalized to $M_0$
as functions of $\mu/M_0$ 
at $\mu_I=T=0$
for nonzero $m_0$. 
The transition from
a constant chiral condensate to a condensate with 
a nonzero wavevector $b$ is first order. 
\begin{figure}[htb]
\begin{center}
\includegraphics[width=0.4\textwidth]{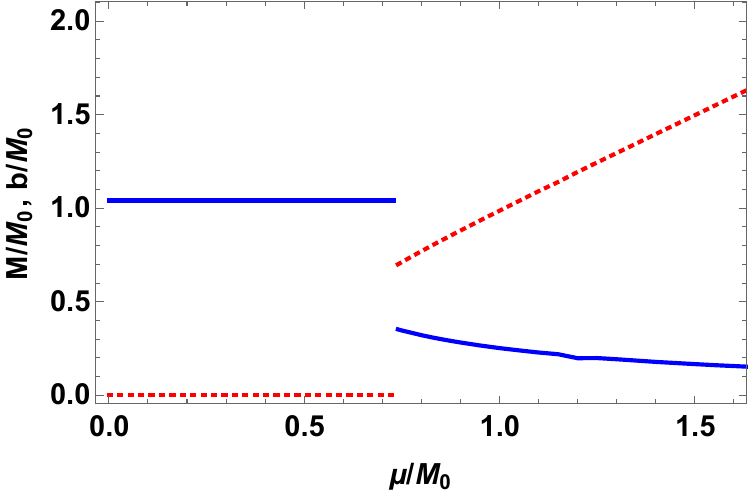}
\end{center}
\caption{(Color online) Normalized magnitude of the quark condensate
$M/M_0$ (blue solid line)
and wavevector $b/M_0$ (red dashed line)
as functions of $\mu/M_0$ at $\mu_I=T=0$ away from the chiral limit.}
\label{larger}
\end{figure}

Since $b=b(\mu)$ is larger than $M=M(\mu)$ in the inhomogeneous phase,
it is clear that dispersion relation for the $u$-quarks 
is that shown in the right panel of Fig. \ref{density1}. This implies that the
energy of a $u$-quark is zero for the finite momentum
$p_{\rm min}=\sqrt{b^2-M^2}$. This is contrast to the $d$-quarks, which are
always gapped with a gap $M+b$.

\subsection{Finite temperature}
\begin{figure}[htb]
\begin{center}
\includegraphics[width=0.4\textwidth]{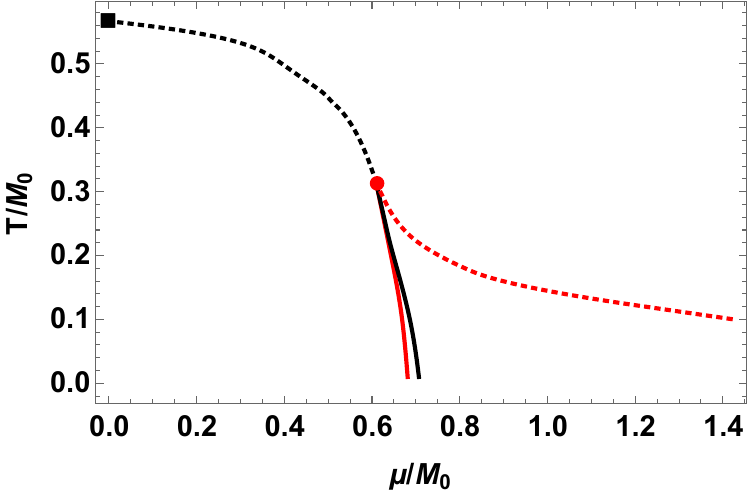}
\end{center}
\caption{(Color online) Phase diagram in the chiral limit. The dashed black and
red lines indicates a
second-order transition, while the solid red line indicates a first-order
transition. The red dot indicates the tricritical point which 
coincides with the Lifschitz point.
The solid black line is the first
order transition in the homogeneous case.}
\label{chiral}
\end{figure}

The complete finite-temperature effective potential is given by the
sum of the vacuum term (\ref{effpot0}) and Eq. (\ref{fint}).
In Fig. \ref{chiral}, we show the phase diagram in the chiral limit.
This phase diagram was first obtained by Ebert et al \cite{symren1}.
The dashed black and red lines indicate second-order transitions,
while the solid red line indicates a first-order transition.
Note that the phase with nonzero
$M$ and $b$ extends to infinity for $T=0$.  
The red dot shows the position of the 
Lifschitz point whose
coordinates in the chiral limit
will be given below. The black solid line indicates the first-order
transition in the homogeneous case. 
In the chiral limit, the tricritical point 
coincides with the Lifschitz point as will be shown below.

\vspace{1cm}
\begin{figure}[htb]
\begin{center}
\includegraphics[width=0.4\textwidth]{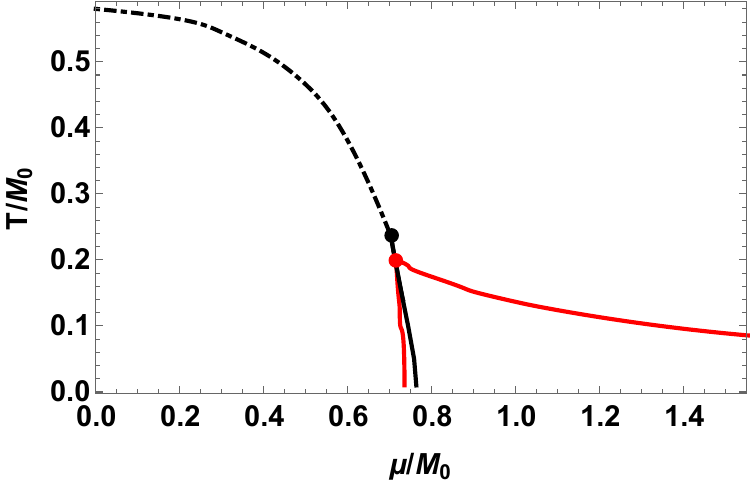}
\end{center}
\caption{(Color online) Phase diagram away from the chiral limit.
The dashed-dotted line is a crossover and the solid red line is a first-order
transition. 
The black dot indicates the critical end point and the red dot
indicates the Lifshitz point, 
and the solid black line is the first
order transition in the homogeneous case.}
\label{physical}
\end{figure}

In Fig. \ref{physical}, we show the phase diagram away from the chiral
limit. 
%Both transitions from and to the symmetric phase are first order. 
Note that the 
position of the critical point (black)
and the Lifschitz point (red) 
do not coincide, in contrast to the result
in the chiral limit.
In the chiral limit, the position of the critical end point and the tricritical 
point can also be found from a Ginzburg-Landau analysis.
We then expand the effective potential in powers of $M$ and derivatives.
In the chiral limit, the first few terms of this expansion are 
\bqa\nonumber
V&=&{N_cM^2\over4G}-2N_cM^2\sumint_{\{P\}}{1\over P^2}
+N_cM^4\sumint_{\{P\}}{1\over P^4}
\\
&&
-{1\over2}N_c(\nabla M)^2
\sumint_{\{P\}}{p^2-3P_0^2\over P^6}\;.
\eqa
%\end{widetext}
We denote by $\beta_1$,
$\beta_2$ and $\beta_3$ the coefficients of $M^2$,
$M^4$ and $(\nabla M)^2$, respectively.
It is easy to show by direct integration over $p$ or by partial
integration, 
that the
the coefficients $\beta_2$ and $\beta_3$ are 
equal. The coefficients are equal also if one uses
momentum cutoff regularization. This is in contrast to three dimensions
where only dimensional regularization \cite{prabal}
or Pauli-Villars regularization \cite{nickel} yield $\beta_2=\beta_3$ 
due to the absence of surface terms.
The tricritical point is given by the condition that the  
quadratic and quartic terms vanish, and the
Lifschitz point is given by the condition that the quadratic and 
gradient terms vanish. The equality of $\beta_2$ and $\beta_3$
implies that the critical point and the Lifschitz point coincide.
The condition that these coefficients vanish implies the coupled
equations
\bqa
\label{beta1}
{1\over8G}-\sumint_{\{P\}}{1\over P^2}&=&0
\;,\\
\sumint_{\{P\}}{1\over P^4}&=&0\;.
\label{beta2}
\eqa
The coefficients $\beta_i$ ($i=1,2,3$)
are all infrared safe since the fermionic
Matsubara frequencies are nonzero. If one separates the
sum-integrals
in a vacuum term and a term that depends on $T$ and $\mu$,
they are both divergent in the infrared, but the divergences cancel
in the sum. The sum-integral $\sumint_{\{P\}}{1\over P^2}$
is also UV divergent and Eq. (\ref{beta1})
needs to be renormalized. 
The sum-integrals appearing in Eqs. (\ref{beta1})--(\ref{beta2})
are calculated in Appendix \ref{app1}.
Using the expression (\ref{addleik}) and making the substitution
${1\over G}\rightarrow Z_{G^{-1}}{1\over G}$,
the renormalized version of Eq. (\ref{beta1})
reads
\bqa\nonumber
\hspace{-3cm}
{1\over8G}-
{1\over2\pi}\left[
\log{\Lambda_{\rm}\over2T}+\gamma_E%-\log2
+{\rm Li}^{\prime}_{\rm -2\epsilon_{\rm IR}}(-e^{-\beta\mu})
%+{\partial{\rm Li}_{\rm -2\epsilon_{\rm IR}}
%(-e^{-\beta\mu})\over\epsilon_{\rm IR}}\bigg|_{\epsilon_{\rm IR}=0}
\right.
\hspace{-3cm}
&& \\ \left. 
\hspace{-1cm}
+{\rm Li}^{\prime}_{\rm -2\epsilon_{\rm IR}}(-e^{\beta\mu})
%+{\partial{\rm Li}_{\rm -2\epsilon_{\rm IR}}
%(-e^{\beta\mu})\over\epsilon_{\rm IR}}
%\bigg|_{\epsilon_{\rm IR}=0}
\right]&=&0
\label{div2222}
\;,
\eqa
where $\Lambda=\Lambda_{\rm UV}$ and $\epsilon=\epsilon_{\rm UV}$.
Using Eq. (\ref{nona}), we can trade $G$ for $M_0$ and 
Eq. (\ref{div2222}) can be written as
%\begin{widetext}
\bqa\nonumber
{1\over2\pi}\left[
\log{M_0\over2T}+\gamma_E%-\log2
%\right. \\ &&\left.
+{\rm Li}^{\prime}_{\rm -2\epsilon_{\rm IR}}(-e^{-\beta\mu})
%+{\partial{\rm Li}_{\rm -2\epsilon_{\rm IR}}
%(-e^{-\beta\mu})\over\epsilon_{\rm IR}}
%\bigg|_{\epsilon_{\rm IR}=0}
\hspace{-1cm}
\right. \\\left.
\hspace{-2cm}
+{\rm Li}^{\prime}_{\rm -2\epsilon_{\rm IR}}(-e^{\beta\mu})
%+{\partial{\rm Li}_{\rm -2\epsilon_{\rm IR}}
%(-e^{\beta\mu})\over\epsilon_{\rm IR}}
%\bigg|_{\epsilon_{\rm IR}=0}
\right]&=&0
\label{firsty}
\;.
\eqa
%\end{widetext}
Using Eq. (\ref{i2}), Eq. (\ref{beta2}) can be conveniently written as
\bqa
{1\over32\pi^2T^3}\left[
\psi(\mbox{$1\over2$}+\mbox{$i\mu\over2\pi T$})
+\psi(\mbox{$1\over2$}-\mbox{$i\mu\over2\pi T$})
\right]&=&0\;.
\label{sgond}
\eqa
The solution to Eqs. (\ref{firsty}) and (\ref{sgond}) gives the
position of the Lifschitz point in the $\mu$--$T$ plane. The solution is
$(\mu/M_0,T/M_0)=(0.6082,0.3183)$ and equals the 
tricritical point
in the chiral limit. The position agrees with 
the numerical result from the phase diagram shown in Fig. \ref{chiral}. 
In the same manner we can find
the critical temperature for the transition at $\mu=0$.
Eq. (\ref{firsty}) reduces to
\bqa
{1\over2\pi}\left[\log{M_0\over\pi T}
+\gamma_E
\right]&=&0\;,
\eqa
whose solution is ${T\over M_0}={e^{\gamma_E}\over\pi}\approx 0.567$.
The point $(0.567,0)$ is marked with a black square
in Fig. \ref{chiral}.

\section{Chiral-density wave versus homogeneous pion condensate}
In this section, we include the possibility of a constant pion 
condensate. 
%A constant pion condensate will form at $T=0$ once the 
%isospin chemical potential exceeds the mass of the pion.

\subsection{Zero temperature}
\begin{figure}[htb]
\begin{center}
\includegraphics[width=0.45\textwidth]{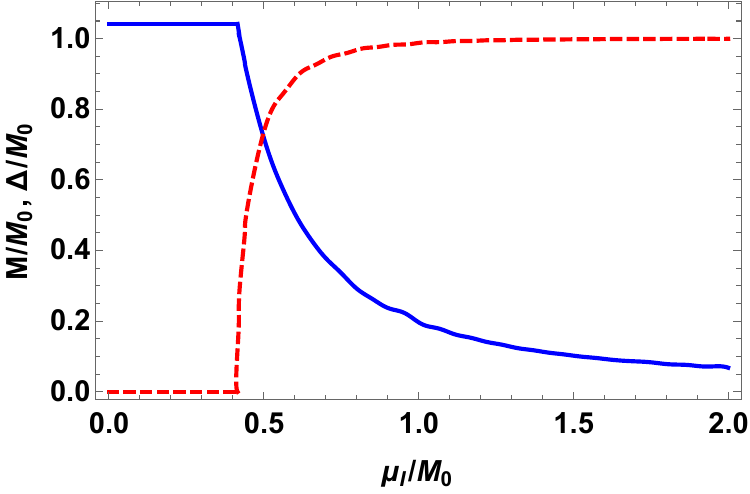}
\end{center}
\caption{(Color online) Normalized quark $M/M_0$
(blue solid line) and pion condensates $\Delta/M_0$
(red dashed line)
as functions of $\mu_I/M_0$ at $\mu=T=0$.}
\label{compgaps}
\end{figure}

In Fig. \ref{compgaps},
we show the normalized quark and pion condensates as functions
of the isospin chemical potential divided by $M_0$
at zero baryon chemical potential and
at zero temperature. For $\mu=0$, the wavevector $b$ vanishes.
The pions condense for $\mu_I\geq\mu_I^c$, where 
$\mu_I^c=m_{\pi}$ is the pion mass in the vacuum phase.
In units of $M_0$, this is approximately $0.42$.
In this phase, the charged pion is a massless Goldstone bosons
associated with the breaking of the $U_{I_3}(1)$ symmetry.
Once the pion condensate starts increasing, 
the quark condensate drops, which can be thought of as a rotation
of the quark condensate into a pion condensate as $\mu_I$ increases.
In the chiral limit, the pion condensate forms for $\mu_I$ infinitesimally
larger than zero and the quark condensate vanishes identically \cite{pionin}.
More generally, in the chiral limit, there is no solution to the gap
equations with $M\neq0$ and $\Delta\neq0$ simultaneously \cite{pionin}.

\begin{figure}[htb]
\begin{center}
\includegraphics[width=0.45\textwidth]{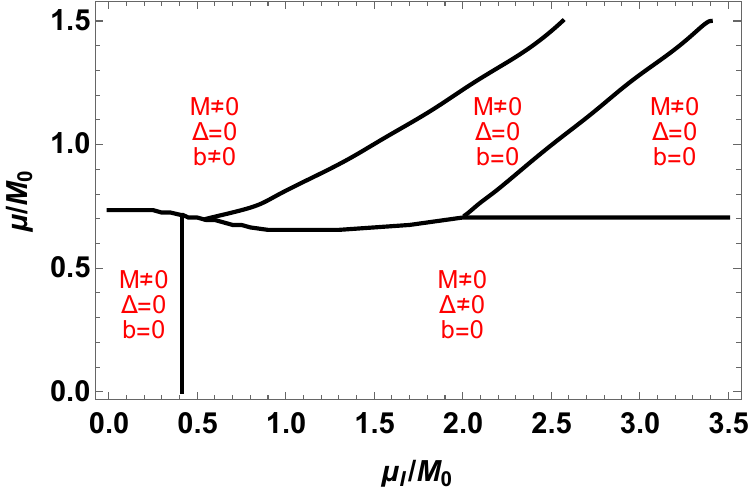}
\end{center}
\caption{(Color online) Phase diagram away from the chiral limit.
%in the $\mu_I$--$\mu$ plane at $T=0$.
See main text for details.}
\label{fullphase}
\end{figure}

In Fig. \ref{fullphase}, we show the 
phase diagram for nonzero quark masses
in the $\mu_I$--$\mu_B$ plane at $T=0$. The values of
$M$, $b$, and $\Delta$ are shown for the different regions.
The transition from the vacuum phase to the phase with a homogeneous  
pion condensate is second order. The other transitions are all 
first order with a jump in the value of $M$ and possibly a jump in the value
of $b$.
This phase diagram generalizes Fig. 5 of Ref. \cite{massive} in which
only constant condensates were considered. The phase with $M\neq0$
and $b\neq0$ for large values of $\mu$ and small values of $\mu_I$
replaces the phase with a constant chiral condensate.
The region in the lower left corner of the
$\mu$--$\mu_I$ plane where $M=M_0$ and $\Delta=b=0$
is the vacuum. In this region it can be shown by taking appropriate
derivatives of the partition function, that physical quantities
are independent of the chemical potentials $\mu$ and $\mu_I$.
This is an example of the silver blaze property \cite{cohen}.
As mentioned above, in the chiral limit, the pion condensate forms
for $\mu_I$ infinitesimally small.
Thus the vacuum phase %with $M\neq0$ and $b=\Delta=0$
reduces to a line along the $\mu$-axis. 
%In this case the 
%critical value $\mu_c$ for an inhomogeneous quark condensate

In the left panel of
Fig. \ref{gapleik}, we show the condensate $M/M_0$ as a function
of $\mu_I/M_0$ for $\mu/M_0=0.9$ in the homogeneouos case, i.e. we do not
allow for a nonzero wavevector $b$. The two transitions are of first order.
In the right panel of Fig. \ref{gapleik}, we show the condensate $M/M_0$ 
and $b/M_0$ as functions
of $\mu_I/M_0$ for $\mu/M_0=0.9$ in the inhomogeneouos case, i.e. we 
allow for a nonzero wavevector $b$. The two transitions are of first order.
This plot corresponds to a horizontal line in Fig. \ref{fullphase}
with $\mu/M_0=0.9$.

\begin{figure}[htb]
\end{figure}

\begin{widetext}

\begin{figure}[htb]
\begin{center}
\includegraphics[width=0.45\textwidth]{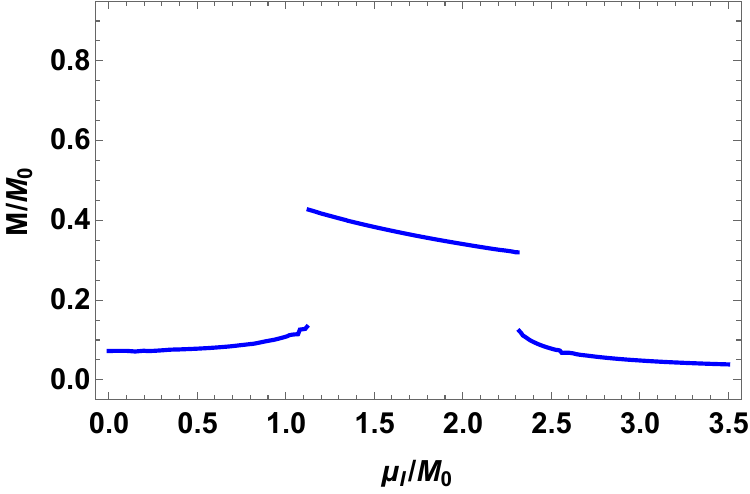}
\includegraphics[width=0.45\textwidth]{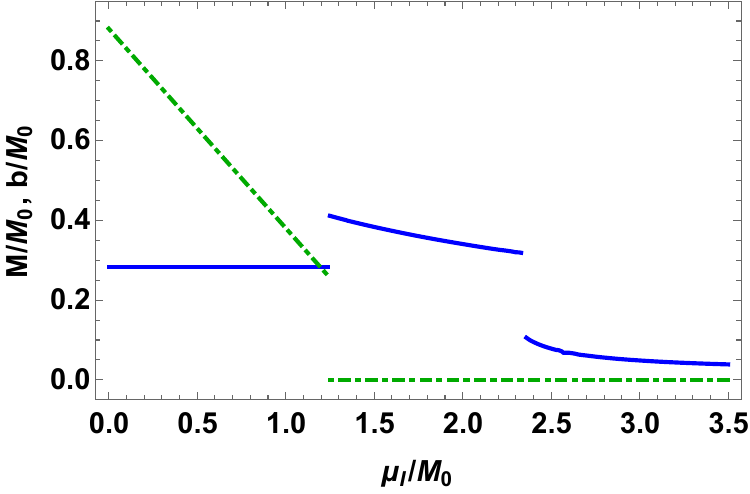}
\end{center}
\caption{(Color online) Left panel:
Normalized quark condensate $M/M_0$ at $T=0$
as function of $\mu_I/M_0$ for $\mu/M_0=0.9$ for $b=0$.
Right panel:
Normalized magnitude of the quark 
condensate $M/M_0$ (blue line) and wavevector $b/M_0$ (green line)
at $T=0$ as functions of $\mu_I/M_0$ for $\mu/M_0=0.9$.
}
\label{gapleik}
\end{figure}

%\begin{figure}[htb]
%\begin{center}
%\includegraphics[width=0.4\textwidth]{TTc.pdf}
%\includegraphics[width=0.45\textwidth]{gapmu09.pdf}
%\end{center}
%\caption{(Color online) Normalized magnitude of the quark 
%condensate $M/M_0$ (blue line) and wavevector $b/M_0$ (green line)
%at $T=0$ as functions of $\mu_I/M_0$ for $\mu/M_0=0.9$.}
%\label{gapleik2}
%\end{figure}
\end{widetext}
\begin{figure}[htb]
\end{figure}

Let us finally discuss the quark and isospin densities in the different
phases. These are given by 
\bqa
%\label{def1}
n_q=-{\partial V^{\rm }\over\partial\mu}\;,
\hspace{1cm}
n_I=-{\partial V^{\rm }\over\partial\mu_I}\;,
\label{def222}
\eqa
where $V^{\rm }=V_0+V_1$ is the full zero-temperature effective potential.
%given by (\ref{fullkli}).
In the phases, where $\Delta=0$, these expressions can be obtained
by differentiation of Eq. (\ref{fullkli}). This yields
\begin{widetext}
\bqa
n_q&=&{N_c\over\pi}\sqrt{(\mu+b^{\prime})^2-M^2}\,\theta(\mu+b^{\prime}-M)
+{N_c\over\pi}\sqrt{(\mu-b^{\prime})^2-M^2}\,\theta(|\mu-b^{\prime}|-M)
{\rm sign}(\mu-b^{\prime})\;,
\\ \nonumber
n_I&=&{N_c\over2\pi}\sqrt{(\mu+b^{\prime})^2-M^2}\,\theta(\mu+b^{\prime}-M)
-{N_{c}\over2\pi}\sqrt{(\mu-b^{\prime})^2-M^2}\,\theta(|\mu-b^{\prime}|-M)
{\rm sign}(\mu-b^{\prime})
\\
&&-{N_c\over\pi}\sqrt{b^{\prime2}-m_0^2}\,\theta(b^{\prime}-m_0)
+{N_c\over\pi}\sqrt{\mbox{$1\over4$}\mu_I^2-m_0^2}
\,\theta(\mbox{$1\over2$}\mu_I-m_0)
\;.
\eqa
\end{widetext}
In the vacuum phase, $b=0$ and so $b^{\prime}=\mbox{$1\over2$}\mu_I$.
Moreover, $M>|\mu\pm\frac{1}{2}\mu_I|$ which implies
that $n_q=n_I=0$. This reflects the 
silver blaze property of the vacuum, namely that its properties
are independent of the chemical potential(s) up to some critical
value(s) above which there is a phase transition.
In the pion-condensed phase, 
%one cannot find simple analytic expressions
the expressions for $n_q$ and $n_I$ follow from (\ref{def222})
and the zero-temperature limit of
Eq. (\ref{total})
(since $b=0$, the subtraction term (\ref{subbi}) vanishes)
\cite{massive}
\bqa
n_q&=&{N_c\over\pi}\int_0^{\infty}\left[
\theta(\mu-E_{\Delta}^+)+\theta(\mu-E_{\Delta}^-)\right]dp\;,
\\ \nonumber
n_I&=&{N_c\over2\pi}\int_0^{\infty}\left[
{E^+\over E_{\Delta}^+}
\theta(E_{\Delta}^+-\mu)
-{E^-\over E_{\Delta}^-}
\theta(E_{\Delta}^--\mu)\right]dp\;.
\\ 
\eqa
Since $E_{\Delta}^{\pm}>\mu$ in this phase, we immediately obtain 
$n_q=0$. The expression for $n_I$ can be found analytically
only in the chiral limit. From Eq. (\ref{vchi}), we find
\bqa
n_I&=&{N_c\over2\pi}\mu_I
%\left[
%\sqrt{\mbox{$1\over4$}\mu_I^2+\Delta^2}-\Delta^2\right]
\;.
\eqa

\subsection{Finite temperature}
\begin{figure}[htb]
\begin{center}
\includegraphics[width=0.45\textwidth]{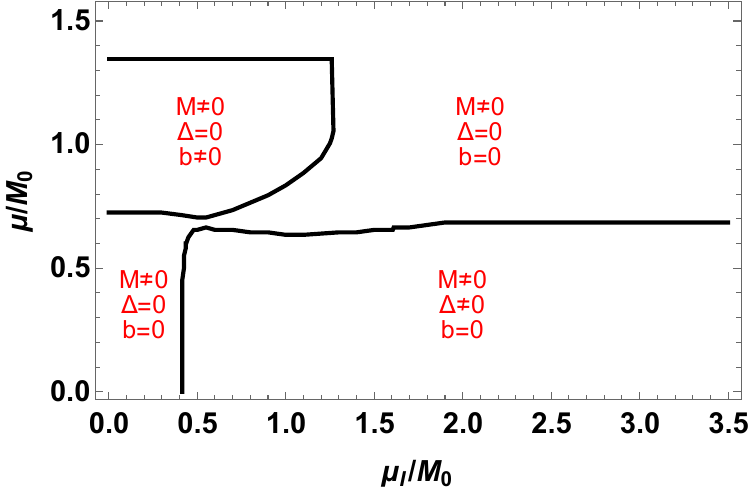}
\end{center}
\caption{(Color online) Phase diagram away from the chiral limit 
%in the $\mu_I$--$\mu$ plane
for $T/M_0=0.1$. See main text for details.}
\label{phase2}
\end{figure}

In Fig. \ref{phase2}, we show the phase diagram for finite quark masses
%in the $\mu_I$--$\mu$--plane
for $T/M_0=0.1$. The inhomogeneous phase now has become an 
island which shrinks as the temperature increases further and eventually
it disappears. The chiral condensate $M$ is continuous through the corridor.
The two phases with $M\neq0$
and $b=\Delta=0$ in the upper right part of Fig. \ref{fullphase}
have now merged into a single phase.
The transitions are all first order. 

\begin{figure}[htb]
\begin{center}
\includegraphics[width=0.45\textwidth]{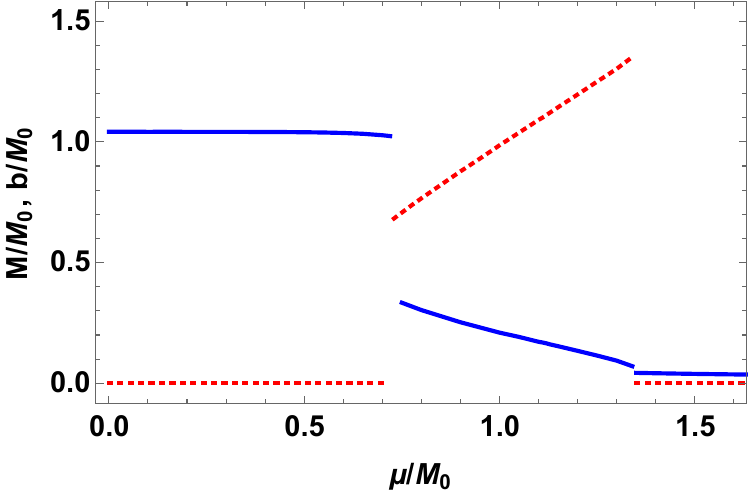}
\end{center}
\caption{(Color online) Normalized quark condensate $M/M_0$
and wavevector $b/M_0$
as functions of $\mu/M_0$
for $\mu_I=0$ and $T/M_0=0.1$.}
\label{gapsie0}
\end{figure}

In Fig. \ref{gapsie0}, we show the
normalized quark condensate $M/M_0$
and wavevector $b/M_0$
as functions of $\mu/M_0$
for $\mu_I=0$ and $T/M_0=0.1$. The two transitions are first order. 

\begin{figure}[htb]
\begin{center}
\includegraphics[width=0.45\textwidth]{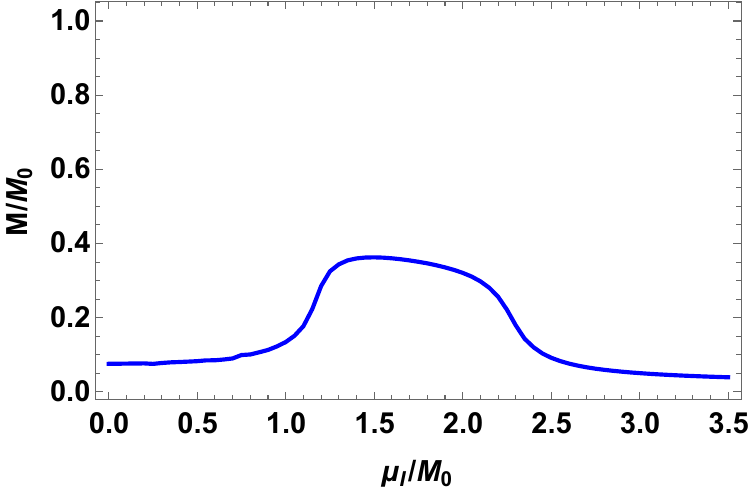}
\end{center}
\caption{(Color online) Normalized chiral condensate $M/M_0$
as a function of $\mu_I/M_0$
for $\mu/M_0=0.9$ and $T/M_0=0.1$ in the homogeneous case, i.e. we do not
allow for nonzero $b$.}
\label{gapsie1}
\end{figure}

In Fig. \ref{gapsie1}, we show the normalized quark condensate
$M/M_0$ as a function of $\mu_I/M_0$
for $\mu/M_0=0.9$ and $T/M_0=0.1$ with the restriction of a constant condensate
i.e. for $b=0$. In contrast to the case at $T=0$, cf. Fig. \ref{gapleik},
$M/M_0$ is continuous.

\begin{figure}%[htb]
\begin{center}
\includegraphics[width=0.45\textwidth]{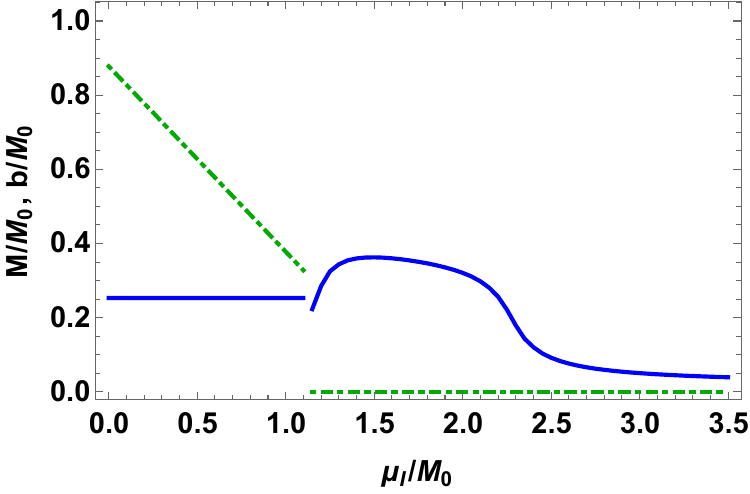}
\end{center}
\caption{(Color online) Normalized chiral condensate $M/M_0$ (blue line)
and wavevector $b/M_0$ (green line)
as functions of $\mu_I/M_0$
for $\mu/M_0=0.9$ and $T/M_0=0.1$.}
\label{gapsie2}
\end{figure}

In Fig. \ref{gapsie2}, we show the normalized quark condensate
$M/M_0$ (blue line) and $b/M_0$ (green line)
as functions of $\mu_I/M_0$
for $\mu/M_0=0.9$ and $T/M_0=0.1$.
$M/M_0$ is discontinuous only for one value of $\mu_I$ showing
that the the phases with $M/M_0\neq0$ and $b=\Delta=0$ have 
merged into a single
phase, cf. the upper right part of Fig. \ref{phase2}.

\section{Summary}
In this paper, we have studied various aspects of the
phase diagram of the NJL model in 1+1 dimensions
in the large-$N_c$ limit as a
function of $T$, $\mu$, and $\mu_I$ using dimensional regularization.
The calculations are done with finite quark masses and generalize
the results of \cite{massive} in which only homogeneous condensates were
considered.

We have also carried out a GL analysis of the tricritical and  
Lifschitz points and derived a set of equations that determine their
position in the $\mu$--$T$ plane. In the chiral limit they coincide, while
they are separated away from it, cf. Figs. \ref{chiral} and \ref{physical}. 
Dimensional regularization proved to
be very useful in their calculation since it can be conveniently
used to regulate
infrared divergences, which cancel in the final result.

In this paper, we restricted ourselves to a constant pion condensate.
The related problem of a constant chiral condensate and an inhomogeneous
pion condensate was considered in Ref. \cite{pionin}.
It would be of interest to extend our calculations to allow
for spatially modulated chiral and pion condensates at the same time.
Eq. (\ref{back2}) would then be replaced by
\bqa
\langle\pi_1\rangle=\Delta \cos(2kz)\;,
\hspace{0.7cm}
\langle\pi_2\rangle=\Delta \sin(2kz)\;,
\eqa
where $k$ is another wavevector.
One complication that arises in the case of an inhomogeneous pion condensate
is that one can no longer find simple analytic expressions for 
the dispersion relations, which means that the problem must be solved numerically in its entirety.
Some work along these lines has been done in the chiral limit 
by Ebert et al \cite{pionin}, 
but a complete mapping of the phase
diagram with nonzero quark masses is still missing.
%The $T=0$ phase diagram 
%in the chiral limit with the restriction to
%constant chiral and pion condensates,
%was worked out in \cite{massive}.

\section*{Acknowledgments}
The authors would like to thank the Niels Bohr International Academy for its 
hospitality.
P.A. would like to acknowledge the research travel support provided by the 
Office of the Dean and the Physics Department at St. Olaf College. P.A. would 
also like to thank Professor Richard Brown at the Computer Science Department 
and Tony Skalski for providing computational support.

%\appendixpage
\appendix
\section{Sum-integrals}
\label{app1}
In this appendix, we evaluate the relevant one-loop sum-integrals
that we need. The sum-integral is defined by
\bqa
\sumint_{\{P\}}&=&
%\left({e^{\gamma_E}\Lambda^2\over4\pi}\right)^{\epsilon}
T\sum_{\{P_0\}}
\int_p
%\int{d^{d}p\over(2\pi)^d}\;,
\label{deffie}
\eqa
{\color{red}
where the integral is defined by
\bqa
\int_p&=&\left({e^{\gamma_E}\Lambda^2\over4\pi}\right)^{\epsilon}
\int{d^{d}p\over(2\pi)^d}\;,
\label{def2}
\eqa}
and $d=1-2\epsilon$,
$P_0=(2n+1)\pi T+i\mu$ are the fermionic Matsubara frequencies,
%with $n=0,\pm1,\pm2...\;$, $\mu$ is the chemical potential,
% for flavor $f$,
and $\Lambda$ is the renormalization scale associated with the
$\overline{\rm MS}$ scheme.

We first consider the sum-integral
\bqa
I_1&=&\sumint_{\{P\}}{1\over P^2}\;.
\eqa
After summing over Matsubara frequencies, we
can write
\bqa\nonumber
%\sumint_{\{P\}}{1\over P^2}&=&
I_1&=&
{1\over2}\int_p{1\over p}\left[1-{1\over e^{\beta(p-\mu)}+1}
-{1\over e^{\beta(p+\mu)}+1}
\right]\;,
\\ &&
\label{i11}
\eqa
The first integral in Eq. (\ref{i11})
which is independent of $\mu$ and $T$
has logarithmic divergences in the infrared and in the 
ultraviolet. The integral
vanishes if the same scale is used in the regularization
of the ultraviolet and infrared divergences \cite{ericaug}. 
If different scales 
are used, the value of the integral is
%%OLD VERSION of A4
\bqa%\nonumber
\int_p{1\over p}&=&{1\over4\pi}\left[
{1\over\epsilon_{\rm UV}}-{1\over\epsilon_{\rm IR}}
+\log{\Lambda_{\rm UV}^2\over\Lambda_{\rm IR}^2}
\right]\;,
\label{vaccont}
\eqa
where the subscripts ${\rm UV}$ and ${\rm IR}$
indicate the different scales.
The second and third integrals in Eq. (\ref{i11})
which depend on $\mu$ and $T$
have logarithmic infrared divergences.
The integrals can also be calculated in 
dimensional regularization and read
\begin{widetext}
\bqa\nonumber
{1\over2}\int_p{1\over p}\left[{1\over e^{\beta(p-\mu)}+1}
+{1\over e^{\beta(p+\mu)}+1}
\right]
&=&-\left({e^{\gamma_E}\Lambda_{\rm IR}^2\over T^2}\right)^{\epsilon_{\rm IR}}
{\Gamma(-2\epsilon_{\rm IR})\over2\sqrt{\pi}\Gamma({1\over2}-\epsilon_{\rm IR})}
\left[
{\rm Li}_{\rm -2\epsilon_{\rm IR}}(-e^{-\beta\mu})
+{\rm Li}_{\rm -2\epsilon_{\rm IR}}(-e^{\beta\mu})
\right]\;,
\\ &&
\eqa
where ${\rm Li}_s(z)$ is the polylogarithmic function with argument $z$
and the subscript ${\rm IR}$ indicates the dimensional regularization
is used to regulate the infrared divergences.
Expanding in powers of $\epsilon_{\rm IR}$ to order $\epsilon_{\rm IR}^0$
yields
\bqa\nonumber
{1\over2}\int_p{1\over p}\left[{1\over e^{\beta(p-\mu)}+1}
+{1\over e^{\beta(p+\mu)}+1}
\right]
&=&-{1\over4\pi}\bigg[
{1\over\epsilon_{\rm IR}}
+\log{\Lambda_{\rm IR}^2\over T^2}+2\gamma_E-2\log2
+2{\rm Li}^{\prime}_{\rm -2\epsilon_{\rm IR}}(-e^{-\beta\mu})
%+2{\partial{\rm Li}_{\rm -2\epsilon_{\rm IR}}
%(-e^{-\beta\mu})\over\epsilon_{\rm IR}}\bigg|_{\epsilon_{\rm IR}=0}
%\right. \\ &&\left.
\\ &&
+2{\rm Li}^{\prime}_{\rm -2\epsilon_{\rm IR}}(-e^{\beta\mu})
%+2{\partial{\rm Li}_{\rm -2\epsilon_{\rm IR}}(-e^{\beta\mu})\over\epsilon_{\rm IR}}
%\bigg|_{\epsilon_{\rm IR}=0}
\bigg]\;,
\label{ftcont}
\eqa
where ${\rm Li}^{\prime}_{\rm -2\epsilon_{\rm IR}}(-e^{\pm\beta\mu})=
{\partial{\rm Li}_{\rm -2\epsilon_{\rm IR}}(-e^{\pm\beta\mu})\over\partial\epsilon_{\rm IR}}
\big|_{\epsilon_{\rm IR}=0}$.
Subtracting Eq. (\ref{ftcont}) from Eq. (\ref{vaccont}), we find
\bqa
%\sumint_{\{P\}}{1\over P^2}
I_1&=&
{1\over4\pi}\left[
{1\over\epsilon_{\rm UV}}
+\log{\Lambda_{\rm UV}^2\over T^2}+2\gamma_E-2\log2
%\right. \\ &&\left.
+2{\rm Li}^{\prime}_{\rm -2\epsilon_{\rm IR}}(-e^{-\beta\mu})
+2{\rm Li}^{\prime}_{\rm -2\epsilon_{\rm IR}}(-e^{\beta\mu})
%+2{\partial{\rm Li}_{\rm -2\epsilon_{\rm IR}}
%(-e^{-\beta\mu})\over\epsilon_{\rm IR}}\bigg|_{\epsilon_{\rm IR}=0}
%+2{\partial{\rm Li}_{\rm -2\epsilon_{\rm IR}}
%(-e^{\beta\mu})\over\epsilon_{\rm IR}}\bigg|_{\epsilon_{\rm IR}=0}
\right]\;.
\label{addleik}
\eqa
\end{widetext}
We note that the poles in $\epsilon_{\rm IR}$ cancel.
Eq. (\ref{addleik}) simplifies in the case $\mu=0$.
Using
${\partial{\rm Li}_{\rm -2\epsilon_{\rm IR}}(-1)\over\partial\epsilon_{\rm IR}}\big|_{\epsilon_{\rm IR}=0}={1\over2}\log{2\over\pi}$, we find
\bqa
I_1&=&
{1\over4\pi}\left[
{1\over\epsilon_{\rm UV}}
+\log{\Lambda_{\rm UV}^2\over\pi^2 T^2}+2\gamma_E\right]\;.
\label{redu}
\eqa
The second sum-integral we need is
\bqa
I_2&=&\sumint_{\{P\}}{1\over P^4}\;.
\eqa
$I_2$ is finite in the infrared as well as in the ultraviolet.
Integration in $d=1$ dimension then yields

\newpage
\bqa\nonumber
I_2&=&{T\over4}\sum_{n=-\infty}^{n=\infty}
{1\over |P_0|^3}
\\ \nonumber
&=&
{1\over32\pi^3T^2}\sum_{n=-\infty}^{n=\infty}
{1\over\big| n+\mbox{$1\over2$}+\mbox{$i\mu\over2\pi T$}\big |^3}
\\ &=& \nonumber
{1\over32\pi^3T^2}\left[
\zeta(3,\mbox{$1\over2$}+\mbox{$i\mu\over2\pi T$})
+\zeta(3,\mbox{$1\over2$}-\mbox{$i\mu\over2\pi T$})
\right]\;,
\\ &&
\label{i2}
\eqa
where $\zeta(n,z)$ is the Hurwitz zeta function.
\section{Vacuum energy for $M=0$, $\Delta\neq0$}
\label{app2}
We next show that the vacuum energy is independent of $b$
in the limit $M\rightarrow0$. We therefore set
$m_0=M=0$ (if $m_0$
is nonzero, so is $M$). The dispersion relation
reduces to $E_{\Delta}^{\pm}=\sqrt{(p\pm b^{\prime})^2+\Delta^2}$.
\begin{widetext}
After integrating over angles, we write the the one-loop cotributions to the
effective potential as 
%\begin{widetext}
$V_{}^{\rm vac}=V_{\rm div}^{\rm vac}+V_{\rm fin}^{\rm vac}$, where
\bqa
V_{\rm div}^{\rm vac}&=&
-{2N_c(e^{\gamma_E}\Lambda^2)^{\epsilon}\over\sqrt{\pi}\Gamma({1\over2}-\epsilon)}
\int_0^{\infty}\sqrt{p^2+\Delta^2}p^{-2\epsilon}\,dp\;,
\\
V_{\rm fin}^{\rm vac}&=&
-{N_c(e^{\gamma_E}\Lambda^2)^{\epsilon}\over\sqrt{\pi}\Gamma({1\over2}-\epsilon)}
\int_0^{\infty}
\left[\sqrt{(p+b^{\prime})^2+\Delta^2}+\sqrt{(p-b^{\prime})^2+\Delta^2}
-2\sqrt{p^2+\Delta^2}\right]p^{-2\epsilon}\,dp
\;.
\eqa
\end{widetext}
Integration gives
\bqa
V_{\rm div}^{\rm vac}&=&{N_c\over2\pi}\left(
{e^{\gamma_E}\Lambda^2\over \Delta^2}\right)^{\epsilon}
\Delta^2\Gamma(-1+\epsilon)\;,
\\ 
V_{\rm fin}^{\rm vac}&=&
-{N_c\over\pi}b^{\prime2}\;,
\eqa
where we have evaluated $V_{\rm fin}^{\rm vac}$ in $d=1$ dimensions.
The term $V_{\rm fin}^{\rm vac}$ is exactly equal to the subtraction term
$f(0,b^{\prime})$ and so $V$ is independent of $b^{\prime}$.
After renormalization and adding the term $f(0,\mbox{$1\over2$}\mu_I)$, we find
\bqa%\nonumber
V&=&{N_c\Delta^2\over4G}
-{N_c\Delta^2\over2\pi}\left[
\log{\Lambda^2\over\Delta^2}+1
\right]
-{N_c\over4\pi}\mu_I^2\;.
%\\ 
\label{vchi}
\eqa
For $\mu_I=0$,
this result is identical to the vacuum energy (\ref{vacci}), which
is a consequence of the fact that the vacuum energy depends on the
quantity $M^2+\Delta^2$.

\bibliography{refs}{}
\bibliographystyle{apsrmp4-1}

\end{document}